%Paper: hep-lat/9408021
%From: meurice@hepmips.physics.uiowa.edu (Yannick Meurice)
%Date: Tue, 30 Aug 94 17:23:36 -0500

%Two mathematica programs are provided after the instruction \end
\input phyzzx.tex
\date{August 25, 1994}
\voffset=3pc
\hoffset=0.45in
%%%%%%%%%%%%%%%%%%%%%%%%%%%%%%%%%%%%%%%%%%%%%%%%%%%%%%%%%%%%%%%%%%%%%%%%%%%%%%%

%%REFERENCES%%%%%%%%%%%%%%%%%%%%%%%%%%%%%%%%%%%%%%%%%%%%%%%%%%%%%%%%%%%%%%%%%%
\def\PL  #1 #2 #3 {{\sl Phys.~Lett.}~{\bf#1} (#3) #2 }
\def\NP  #1 #2 #3 {{\sl Nucl.~Phys.}~{\bf#1} (#3) #2 }
\def\PR  #1 #2 #3 {{\sl Phys.~Rev.}~{\bf#1} (#3) #2 }
\def\PRD #1 #2 #3 {{\sl Phys.~Rev.~D} {\bf#1} (#3) #2 }
\def\PRB #1 #2 #3 {{\sl Phys.~Rev.~B} {\bf#1} (#3) #2 }
\def\PP  #1 #2 #3 {{\sl Phys.~Rep.}~{\bf#1} (#3) #2 }
\def\MPL #1 #2 #3 {{\sl Mod.~Phys.~Lett.}~{\bf#1} (#3) #2 }
\def\CMP #1 #2 #3 {{\sl Comm.~Math.~Phys.}~{\bf#1} (#3) #2 }
\def\PRL #1 #2 #3 {{\sl Phys.~Rev.~Lett.}~{\bf#1} (#3) #2 }
\def\TMP  #1 #2 #3 {{\sl Theor.~Math.~Phys.}~{\bf#1} (#3) #2 }
\def\JMP  #1 #2 #3 {{\sl Jour.~Math.~Phys.}~{\bf#1} (#3) #2 }
\def\JSP  #1 #2 #3 {{\sl Jour.~Stat.~Phys.}~{\bf#1} (#3) #2 }
\def\IJ  #1 #2 #3 {{\sl Int.~Jou.~Mod.~Phys.}~{\bf#1} (#3) #2 }
%%%%%%%%%%%%%%%%%%%%%%%%%%%%%%%%%%%%%%%%%%%%%%%%%%%%%%%%%%%%%%%%%%%%%%%%
%
\REF\dyson{F. Dyson, \CMP 12 91 1969 ; G. Baker, \PRB 5 2622 1972 . }
\REF\wilson{K.~Wilson, \PRB 4 3185 1971 ; K. Wilson
and J. Kogut, {\sl Phys. Rep.}
{\bf 12} (1974) 75  .}
\REF\sinai{P. ~Bleher and Y. ~Sinai, \CMP 45 247 1975 ; P.~Collet and
J. P. ~Eckmann,\CMP 55 67 1977 and {\it
Lecture Notes in Physics} {\bf 74} (1978) ;
K. Gawedzki
and A. Kupiainen, Les Houches 1985, K. Osterwalder and R. Stora, Editors .}
%
%\REF\epsi{P. Collet, J.-P. Eckmann, and B.
%Hirsbrunner, {\it Phys. Lett.} {\bf 71B}, 385 (1977). }
%
%\REF\lecture{P. Collet and J.-P. Eckmann,  }
%
%
\REF\num{Y. Meurice, G. Ordaz and V.G.J. Rodgers, \JSP 77 0000 1994 (in press)}
\REF\app{Y. Meurice, G. Ordaz, U. of Iowa Preprint 94-15}
\REF\conje{Y. Meurice, \JMP 35  769 1994 .}
%
%\REF\polyakov{A. ~Polyakov, \PL 82B 247  1979  and  \PL 103B 211 1981 . }
%
%\REF\parisi{G. Parisi, {\it Statistical Field Theory} (Addison-Wesley,
%New-York, 1988) and references therein. }
%
%\REF\dis{There is a large amount of literature on this subject. References can
%be found in {\it Phase Transitions, Cargese 1980}, M. Levy, J.C. Le
%Guillou and J. Zinn-Justin Editors, (Plenum Press, New York, 1982). }
%
%\REF\confl{Z. Roskies \PRB 24 5305 1981 ; J. Adler, M. Moshe and
%V. Privman \PRB 26 3958 1982 ; G. Nickel and M. Dixon
%\PRB 26 3965 1982 .}
%
\REF\missarov{E. Lerner and M. Missarov, \TMP 78 177 1989 .}
\REF\rw{J.L.Lucio and Y. Meurice,
\MPL 6 1199 1991 ; Y.Meurice,
{\sl Phys.~Lett.}~{\bf 265B} (1991) 377 . }
\REF\taibleson{M. Taibleson, {\it Fourier Analysis on Local Fields}, Princeton
University Press.}
\REF\serre{J.P. Serre, {\it A Course in Arithmetics}, (Springer, New York,
1973). }
\REF\brydges{K. Symanzik, in {\it Local Quantum Theory}, R. Jost Editor,
Academic Press, New York, 1969 ;
D.~Brydges, J.~Fr\"ohlich and T. ~Spencer, {\sl Comm. Math. Phys. }
{\bf 83} (1982) 123 ;
C. Itzykson and J.M Drouffe, {\it Statistical Field Theory}
, Cambridge University Press, Cambridge, 1989.}
%%%%%%%%%%%%%%%%%%%%%%%%%%%%%%%%%%%%%%%%%%%%%%%%%%%%%%%%%%%%%%%%%%%%%%%%%%
\Pubnum={U.Iowa 94-14\cr
hep-lat/9408021}
\titlepage
\title{ The High-Temperature Expansion of the Hierarchical Ising Model:
{}From Poincar\'e Symmetry to an Algebraic Algorithm}
\author{Y. Meurice }
\address{Department of Physics and Astronomy, University of Iowa,
Iowa City, Iowa 52242, USA}
\vfil
\abstract
We show that the hierarchical model at finite volume has a symmetry
group which can be decomposed into rotations and
translations as the familiar Poincar\'e groups.
Using these symmetries, we show
that the intricate sums appearing in the calculation of the high-temperature
expansion of the magnetic susceptibility
can be performed, at least up to the fourth order,
using elementary algebraic manipulations which can be implemented
with a computer. These symmetries appear more clearly if we use the 2-adic
fractions to label the sites. We then apply
the new algebraic methods to the calculation
of quantities having a random walk
interpretation. In particular, we show that the probability of returning
at the starting point after $m$ steps has poles at $D=-2,-4,....-2m$ ,
where $D$ is a free parameter playing a role similar to the dimensionality
in nearest neighbor models.

\endpage

\chapter{Introduction}

The hierarchical model\refmark\dyson
played an important role in the development
of the ideas of the renormalization group,\refmark\wilson
because the renormalization
group transformation for this model reduces to a simple integral equation
which
has been studied in great detail.\refmark{\sinai }
In a recent paper,\refmark{\num} we calculated numerically the magnetic
susceptibility
of the hierarchical
Ising model as a function of the temperature and for two values
the parameter used
in the $\epsilon$-expansion. Calculations have been carried
with up to $2^{18}$ sites.
We found that the numerical data can be fitted precisely
with a simple power law of
the form $(1- \beta /\beta _0 )^{- g}$
in the {\it whole} high-temperature region, i.e for $\beta \in [0,\beta _c]$.
%It seems that in the critical region, this result
%is an artifact of the finite volume. On the other hand, in the
%high-temperature region, the dependence on the volume appears to be
%small for more than $2^{13}$ sites.
As a consequence, it is possible to obtain a simple approximate formula for
the high-temperature coefficients of the susceptibility in terms of $g$ and
$\beta _0$.
This approximate formula implies approximate relations\refmark{\num,\app} among
the coefficients which deserve a systematic investigation.
Unfortunately, due to the non-locality of the hierarchical model, the
calculation
of the high-temperature expansion is quite non-trivial even for the very
first orders.

The goal of this article is to explain in a pedestrian way how the symmetries
of the hierarchical model can be used to reduce the
convoluted sums appearing in the high temperature expansion,
at least up to order 4 in $\beta $,
to simple algebraic manipulations which can be
easily implemented with a computer.
In section 2, we show that even at finite volume, the hamiltonian of the
hierarchical model is
invariant under a group of transformation which can be decomposed into
translations and rotations as the familiar Poincar\'e groups.

One important reason to work at finite volume is that high-temperature
coefficients of arbitrary order can be calculated with an arbitrary precision
using a computer, if the number of sites
is not too large. This can be done by introducing a $\beta$
expansion within the exact method of integration described in Ref. [\num].
Since it is very easy to make combinatoric mistakes in calculating
high-temperature coefficients, formulas valid for arbitrary volume allow
unambiguous checks if we can compare them with sufficiently precise
numerical results obtained with a finite number of site.
A numerical program performing this calculation is provided in Appendix 2.
This program can be used to check the validity of the analytic results
presented hereafter with at least 10 significant figures, when the
number of sites does not exceed $2^9$.
More fundamentally, the renormalization group approach requires techniques of
integration at finite volume and the results obtained here
can be used in this context.

In section 3 , we use extensively the translation invariance to calculate
the first four coefficients of the magnetic susceptibility. The final result
is that each of these coefficients can be written a sequence of products and
Fourier transforms. In section 4, we show that these sequences of
operations can be reduced
to simple algebraic manipulations which can be implemented on a computer.
A ready-to-use Mathematica program is provided in Appendix 1.
The methods presented here  can be used for higher order calculations,
however it is not clear that the calculation of coefficients of
an arbitrary order can be completely reduced to simple algebraic
manipulations. This issue is briefly discussed  in the conclusions.

The results of section 2 and 4  could have been
derived more elegantly if we had
labeled the sites using 2-adic fractions as in Ref. [\missarov].
This is explained in section 5 which, unlike section 2, assumes some
familiarity with the mathematics involved in this reformulation.
However, section 5 is not essential to the understanding of
the other sections.

The results obtained in section 4 take a more compact form
in the infinite volume limit.
In section 6, we give the explicit form
of each terms appearing in the calculation of the first four
coefficients in this limit. These expressions have poles
at unphysical values of $D$, a parameter controlling the strength
of the bilinear couplings among the spins (see section 2), and playing
a role similar to the dimension in nearest neighbor models.
By unphysical values of $D$, we mean values such that
changing the sign of one spin in one of the configuration where
all the spins are aligned causes an infinite increase in energy.
Incidently, all the quantities calculated
in that section have a random walk interpretation,
provided that a proper normalization is used. In general, the calculation
the $m$-th coefficient of the high temperature expansion involves the
calculation of the probability of returning at the starting point after $m$
steps and the probability of visiting less than two sites in $m$ steps.
In the infinite volume limit, these quantities can be calculated for
arbitrary $m$. Their poles are located at $D=-2,-4,....,-2m$ and
$D=-{{2m}\over {m-1}}$ respectively.

\chapter{The Hierarchical Ising Model and its Poincar\'e Group}

In this section we describe the hierarchical Ising  model and its invariance
under a group of transformation which can be decomposed into translations
and rotations in a way similar to the Poincar\'e group used in relativistic
quantum
field theory.
Hierarchical models\refmark\dyson
are specified by a non-local hamiltonian bilinear
in the spin variables and a local measure of integration
which will not
be specified in this section. The specific choice of an Ising
measure where the spins take only the values $\pm 1$ will only
be made in the next
sections.
We first recall the form of the non-local hamiltonian.

The hierarchical models considered here require the number of sites to be
$2^n$.
For convenience, we label the sites with $n$
indices $x_n ..... x_1$, each index
being 0 or 1 .
In order to give a concrete meaning to
this notation, one can divide
the $2^n$ sites into two boxes, each containing $2^{n-1}$
sites. If $x_n =0$, the site is
the first box, if $x_n=1$, the site is in the second box.
Repeating this procedure $n$ times
(for the two boxes, their respective two sub-boxes, etc.),
we obtain an unambiguous labeling for each of the sites.
In the following, we often use the symbol $x$ as a short notation
for the sequence $x_n ..... x_1$.

The hamiltonian of the hierarchical model reads
$$H=-{1 \over 2} \sum\limits_{l=1}^{n}({c \over 4})^l
\sum\limits_{x_n,...,x_{l+1}}\ (\ \sum\limits_{x_l,....,x_1} \sigma
_{(x_n,....,x_1)} )^2\ . \ \eqno(2.1) $$
The model has a free parameter $c$ for which we shall use the
parametrization
$$c=2^{1-{2\over D}}.\eqno(2.2)$$
The parameter $D$ plays a role similar to the dimension in nearest
neighbor models.
The parameter of the epsilon-expansion can be defined as
$\epsilon = 4-D$.
When $D\geq 4$, the model has a trivial continuum limit.\refmark\sinai
When $D\leq 2$, the model does not have a phase transition at finite
temperature.\refmark\dyson
These two rigorous results can be understood heuristically
in terms of the the self-intersection properties of the random walk
associated with $H$, by noticing that the Haussdorff dimension
of this random walk is $2/D$.\refmark\rw

For a given $l$ in Eq. (2.1), the first sum can be interpreted as
a sum over boxes of size $2^l$, while the second sum is
over the spins inside
each of these boxes. Given two distinct sites $x$ and $y$, the corresponding
spins have an interaction whenever $l$ is large enough to have $x$ and
$y$ within the same box. It is clear that if $x$ and $y$ belong to the same
box of size $2^l$, they also belong to the same boxes of larger size,
until the maximal size is reached. Since the respective contributions
to the interaction are $({c\over 4})^l$ in Eq. (2.1), the total interaction
can be expressed as a truncated geometrical series.
In order to describe quantitatively this situation, we
define a function $v(x,y)$ which indicates
the ``level'' $l$ at which $x$ and $y$ start to differ.
More precisely, if $x$ and $y$ are distinct,
$v(x,y)=l$ when $x_m=y_m$ for all $m$ such that
$n\geq m>l>0$
$and$ $x_l\neq y_l$. At coinciding arguments, we define $v(x,x)=0$.
The hamiltonian function $H$ can then be rewritten as
\def\bk{{\bf K}}
\def\bkx{{\bf K}_{xy}}
\def\cov{{c \over 4}}
$$H=-{1\over 2}(\sum _{x,y} \bkx \sigma _x \sigma _y + {\bf L}\sum _x
\sigma _x ^2 )\eqno(2.3)$$
where
$$\bkx = \cases{ ((\cov )^{v(x,y)}-(\cov )^{n+1})(1-\cov )^{-1}
\ {\rm if}\ x \neq y \cr
0 \ \ \ \ \ \ \ \ \ \ {\rm if }\ x=y \cr }\eqno(2.4)$$
and
$${\bf L}=((\cov )-(\cov )^{n+1})(1-\cov )^{-1}\eqno(2.5)$$

As made clear by the above equations, the strength
of the interaction
between two spins $\sigma _x$ and $\sigma _y$ depends only on the
value of $v(x,y)$. Consequently, the invariance of $v(x,y)$ under a group
of transformation (see Eq. (2.10))
implies the invariance of $H$ under corresponding
transformations, in a way which will be made precise at the end
of this section. The invariance of $v(x,y)$ under a group of
transformation appears rather clearly in the reformulation of the hierarchical
model proposed in Ref. [\missarov ], where the sites are associated
with 2-adic fractions which can be transformed into each other using
addition and multiplication. However, it is possible to obtain
the results that will be needed later using only a few basic results
in arithmetic $modulo\ 2^n$. The reader familiar with the
2-adic numbers will notice that since the 2-adic integers can be
defined\refmark\serre as the projective limit of the integers $modulo
\ 2^n$, the finite volume
presentation given here is implicit in the formulation
of Ref. [\missarov].

In order to describe the invariance of $v(x,y)$, we shall
associate with the sequence of 0's  and 1's $x_n ..... x_1$,
a rational number of the form
\def\frac{(2.6)}
$$x=\sum_{m=1}^n x_m 2^{-m}\eqno\frac $$
Since this defines a one-to-one correspondence, we shall use the same
symbol $x$ for the sequence of 0 and 1 and the rational number.
We can now define additive and multiplicative operations which
will be used to  obtain a group of transformation.

If two numbers $x$ and $y$ have the form given in Eq. $\frac $,
then $x+y$ can also be written
as a rational number where the denominator is $2^m$ with $m\leq n$.
If we drop the integer part of $x+y$, we obtain a rational number
having the form of Eq. $\frac$.
Equivalently, we can write $x=q/2^n$ and $y=r/2^n$ with $q$ and $r$ integers
between 0 and $2^n -1$ and add $q$ and $r$ $modulo$ $2^n$.
Since the integer $modulo\ 2^n$ form an additive group,
the set of fractions associated
with the sites form a group for the
addition modulo 1. In order to clarify the ideas,
let us give an example: the additive
inverse of $2^{-n}$ $modulo$ 1 can be written as
$2^{-n} + 2^{-n+1} + ..... + 2^{-1}$.
This makes clear that the addition $modulo$ 1 is distinct
from the addition obtained by adding each of
the indices ($x_m+y_m$) $modulo$ 2.
Furthermore, the odd integers $modulo$ $2^n$ form a multiplicative
group. We can pick a canonical form for the representatives
of such integers as
\def\un{(2.7)}
$$u=1+2z\eqno\un $$
where $z$ is a positive integer between 0 and $2^{n-1} -1$.
Obviously, if $x$ has the form $\frac$, then $ux$ has
also the form $\frac$ after discarding its integer part.

We are now in position to define a group of transformation
acting on the fractions associated with the sites. If $x$ and $a$
have the form of Eq. $\frac$ and $u$ has the form of Eq. $\un$,
we define a transformation of $x$ depending on $a$ and $u$ and denoted
$x[u,a]$ which reads
$$x[u,a]=ux+a \ ,\eqno(2.8)$$
where the r.h.s is understood $modulo \ 1$.
It is clear that $u=1,a=0$ gives the identity transformation.
{}From the composition law $(x[u,a])[v,b]=x[uv,va+b]$, one sees that
an inverse transformation is obtained for $v=u^{-1}$ and $b=-u^{-1}a$.
Consequently these transformations form a (non-abelian) group.

We can
interpret $x[0,a]$ as a
translation and $x[u,0]$ as a rotation like in
the usual Poincar\'e groups. In that sense, this is a ``global''
group of transformation. This is in contrast with the symmetries
noted by Dyson\refmark\dyson which consists in interchanging
$x_m.....x_{l+1} 1 x_{l-1}......x_1$ and $x_m.....x_{l+1} 0 x_{l-1}......x_1$
``locally''.

In the following, we shall frequently use the fact that $x[u,a]$ with
$u$
and $a $ fixed is a
one-to-one map. The proof of this statement is simple. If $ux+a=uy+a\
modulo \ 1$, then $u(x-y)=0\ modulo \ 1$, writing $x-y$ as $q/2^n$, we
obtain that $q=0\ modulo \ 2^n$ and consequently, $x=y\ modulo \ 1$.

We can now make a connection between $v(x,y)$ and $x-y$ through
the following result:
\def\ifo{(2.9)}
$$v(x,y)=l\ \Longleftrightarrow \ x-y=q/2^l \ modulo\ 1 \
{\rm for\ some\ odd\ integer} \ q \ \eqno\ifo .$$
In order to prove this, we write
$x=r/2^n$ and $y=s/2^n$ with $r$ and $s$ integers
between 0 and $2^n -1$ and we see that r.h.s of the equivalence relation
means that $r$ and $s$ are equal $modulo$ $2^{n-l}$ but differ
$modulo$ $2^{n-l+1}$. Returning to the canonical form of Eq. $\frac $,
we see that is equivalent to $v(x,y)=l$.
Noticing that $x-y$ is translation invariant and that the product
of odd integers is still an odd integer, we obtain the invariance
of $v(x,y)$ under the transformations (2.8) in the following way:
$$v(x[u,a],y[u,a])=v(x,y)\ .\ \eqno(2.10)$$
Since $\bkx $ depends only on $v(x,y)$, this immediately implies that
$$\bkx = {\bf K}_{x[u,a] y[u,a]} \ .\eqno(2.11)$$

It is time to state the main result of this section: {\it The hamiltonian
H  defined
by  Eq.  (2.1) is  invariant under  the  transformation}
$$\sigma _x \longrightarrow \sigma_{x[u,a]}\eqno(2.12)$$
\noindent
$for \ any \ a \ as \  in \ Eq. \ \frac \ and \ any \ u \ as \ in \ Eq.\
\un$.

\noindent
This follows from the fact that $x[u,a]$ defines one-to-one map, consequently
we can rewrite the transformed hamiltonian in the following way
$$\sum _{x,y}{\bf K}_{xy}\sigma_{x[u,a]}
\sigma _{y[u,a]} = \sum _{x,y}{\bf K}_{x[u^{-1},-u^{-1}a]y[u^{-1},-u^{-1}a]}
\sigma _x \sigma _y \eqno(2.13)$$
and then use the invariance of $\bkx $ shown in Eq. (2.11).

As is well-known, the invariance of $H$ under a group of transformations
implies identities for the correlation functions. In order to define
these, we need to introduce a local measure of integration. In the following
sections, we shall consider the case of an Ising measure and define
$$ Z=\sum_{\{ \sigma = \pm 1 \} }
e ^{-\beta H}\eqno(2.14)$$
and
$$<\sigma _{x}\sigma _{y}>=
Z^{-1} \sum_{\{ \sigma  = \pm 1 \} }
\sigma _{x}\sigma _{y} e ^{-\beta H}\ ,\eqno(2.15)$$
We can then prove that
$$<\sigma_{x[u,a]}\sigma _{y[u,a]}>\ = \ <\sigma _x \sigma _y\ >\eqno(2.16)$$
This follows from the fact
that we can always relabel the integration variables
according to Eq. (2.12.) and then use the invariance of $H$. This argument
is clearly independent of the choice of the local measure.

\chapter{The High-Temperature Expansion of the Hierarchical Ising Model}

We now proceed to the calculation of the magnetic susceptibility per
site for the hierarchical Ising model.
We define the magnetic susceptibility per site as
$$\chi_n(\beta)\ = \ {1\over {2^n}}<\ (\sum _x \sigma _x )^2 \ > \eqno(3.1)$$
Using translation invariance, we find that
$$\chi_n(\beta)\ =\ <\ \sigma_0(\sum_x \sigma _x)\ > \ ,\eqno (3.2)$$
the choice of 0 being arbitrary. In the case of the Ising model we
can use the fact that $\sigma_0^2 =1$ and write
$$\chi_n(\beta)\ =\ 1\ + \ \sum_{x;x\neq 0}
<\ \sigma_0 \sigma _x\ > \ ,\eqno (3.3)$$
\def\sox{<\ \sigma_0 \sigma _x\ >}
\def\bky{{\bf K}_{yz}}
Using the high-temperature expansion
of $ <\ \sigma_0 \sigma _x\ >$, we can calculate the coefficients of
the expansion
$$\chi_n (\beta)=1\ + \ b_{1,n}\beta \ + \ b_{2,n}
\beta ^2 \ + \ ... \eqno(3.4)$$
A brief review of the standard methods
used to perform such a calculation with arbitrary long-range Ising
models is given at the beginning of section II of Ref. [\conje]. For the
sake of briefness, we shall only recall the final result: $\sox$ can be written
as
the ratio of two expansions in $tanh(\beta \bky) $. The terms of each
of them correspond
to graphs where a given link $y-z$ can only appear once (in which case
we have a have a factor $tanh(\beta \bky) $) or zero times (in which case we
have a factor 1). At the denominator, all the sites must be visited an
even number of times. At the numerator,
all the sites must be visited an even number of times except for 0 and $x$
which are visited an odd number of times.
This provide an expansion in $tanh(\beta \bky) $ which will
require further expansion in order to be recast in the form (3.4).
It has to be noted that for hierarchical models, the use of Feynman
diagrams in the recursion formula can be reduced to a purely combinatoric
problem (no sums or integrals), while the high-temperature expansion
requires the evaluation of nested sums - a situation which somehow
the opposite of what we encounter for nearest neighbor models.
We now proceed order by order to the calculation of the $b_{m,n}$ for
$m=1,2,3,4$ and $n$ arbitrary.

The first order contributions to $\sox $ comes from high-temperature
graphs with one
link joining 0 to $x$. The contribution of these graphs is
$$\sum _x tanh(\beta {\bf K}_{0x} )\ .\eqno(3.5)$$
Note that $\bkx $ is zero when $x=y$ and we do not need to subtract
the contribution with $x=0$. We thus obtain
$$b_{1,n}=\sum_x {\bf K}_{0x} \ . \eqno(3.6)$$
Proceeding similarly with graphs with two links, we obtain
$$b_{2,n}=\sum_{y,x \atop x\neq0}\bk _{0y}K_{yx} \eqno(3.7)$$
At order three, we need to consider the $\beta^3 $ terms of Eq. (3.5)
in addition to the lowest order contributions from the graphs with
three links. The result is
$$b_{3,n}=\sum_{y,z,x \atop z\neq 0, x\neq y,0}\bk _{0y}
\bk_{yz}\bk_{zx}\ - \ {1\over 3}\sum _x (\bk _{0x})^3 \ \eqno(3.8)$$
Up to now, we did not have to take into account the contributions from
the denominator because the first non-trivial contributions
to $Z$ come from triangles which at lowest order give order $\beta ^3$ terms.
Since $\sox$ yields contributions which are at least of order $\beta $
when $x\neq 0$, the denominator has to be taken into account at order 4.
These contributions cancel the lowest order of the graphs with
a triangle at one or the other
end and all which is left are minus the graphs where the link $0-x$
is also occupied by one side of a triangle (since such a graphs do not
appear at the numerator). Adding to this the lowest order contribution from the
graphs
joining 5 distinct sites and the order $\beta ^4 $ of the graphs with
two links, we obtain
$$\eqalign{b_{4,n}=&\sum _{y,z,w,x \atop all\ distinct}
\bk_{0y}\bk_{yz}\bk_{zw}\bk_{wx} \ - \sum_{xy}(\bk_{0x})^2
\bk_{0y}\bk_{xy} \ \cr &- \ {1\over 3}\sum_{x,y}
(\bk_{0y})^3\bk_{yx}-  \ {1\over 3}\sum_{x,y}\bk_{0y}(\bk_{yx})^3
\ +\ {2\over 3} \sum_x (\bk_{0x})^4 }\eqno(3.9) $$
Note that we have taken into account that $\bkx ={\bf K}_{yx}$.

{}From the translation invariance of $\bkx $, it is clear that
$\bkx $ is a function of $x-y$ only. Consequently, we can write
$$\bkx \ = {1 \over 2^n}\sum _{k=0}^{2^n -1} \widehat{ \bk} (k)
e^{-i2\pi k(x-y)} \ . \eqno(3.10)$$
\def\wkn{\widehat{{\bf K}}}
\def\wko{\widehat{{\bf K}}(0)}
\def\wkt{\widehat{{\bf K}^2}(0)}
\def\wkth{\widehat{{\bf K}^3}(0)}
\def\wkf{\widehat{{\bf K}^4}(0)}
The explicit form of
$\wkn (k)$ will be given in Eq. (4.12).
Our task in now to rewrite the expressions for the $b_{m,n}$ obtained
in the previous section as unrestricted sums over $y,\ z,\ ... x$
with appropriate compensating terms, plug the expansion (4.1) in them
and use the well-known results
$$\sum_z\bk_{xz} {\bf G}_{zy}=
{1 \over 2^n}\sum _{k=0}^{2^n -1} \widehat{ \bk} (k)\widehat{{\bf G}}(k)
e^{-i{2\pi}k(x-y)} \ . \eqno(3.11)$$
and
$${1 \over 2^n}\sum _{q=0}^{2^n -1} e^{-i{2\pi\over 2^n}kq} \
= \ \delta_{k,0} . \eqno(3.12)$$
Proceeding this way, we obtain
$$b_{1,n}=\wko \ \eqno(3.13)$$

$$b_{2,n}=(\wko )^2-\wkt  \ \eqno(3.14)$$

$$b_{3,n}=(\wko )^3-2\wkn (0) \wkt+{2\over 3}\wkth -(\widehat{\wkn ^3})(0)
\eqno(3.15)$$

$$\eqalign{b_{4,n}=&{(\wko)^4} + {4 \over 3}\wko \wkth -{4\over 3} \wkf
+2 (\widehat{\widehat{{\bf K}^2}\wkn^2})(0)\cr &-3\wkt(\wko)^2
+3(\wkt)^2
-\widehat{\wkn ^4}(0)-2\wko (\widehat{(\wkn) ^3})(0)}\eqno(3.16)
$$

Note that expressions of the same form but with different coefficients
would be obtained if instead of using a Ising measure, we had
an arbitrary measure. This can be seen in general by using the
random path representation.\refmark\brydges

\chapter{Explicit Expressions for the High-Temperature Coefficients}

Our next task consists in calculating explicitly the above expressions.
For this purpose, we first introduce the function $N(k)$
defined over
the integers $modulo$ $2^n$:
$$ N(k) = \cases{ 2^{-l} & \  {\rm if}\ $ k$ can be divided by
\ $2^{l}$ \
 {\rm but\ not\ by} \  $2^{l+1}$ \cr
2^{-n} & \ {\rm if} \ $ k=0 \ modulo\ 2^n $} \eqno(4.1)$$
This function will appear in $\wkn (k)$.
We now introduce short notations for some functions of $x$
$$2^{-sv(x,0)}\longrightarrow {\cal X}_s$$
$$\delta_{x,0}\longrightarrow \delta\eqno(4.2)$$
and for some functions of $k$
$$(N(k))^{{2\over D}s} \longrightarrow {\cal K}_s$$
$$\delta_{k0}\longrightarrow \Delta \eqno(4.3)$$
In addition, we also define
$$P(1+a)={1\over {1-2^{-(1+a)}}}\eqno(4.4)$$

$$G(1+a)={{1-2^a}\over {1-2^{-(1+a)}}}\eqno(4.5)$$

$$Q(1+a)=2^{-(n+1)(a+1)}\eqno(4.6)$$
With these notations,
\def\tov{{2\over D}}
$${\bf K}_{x0}\longrightarrow P(1+\tov)({\cal X}_{1+\tov}-Q(1+\tov){\cal X}_0
-(1-Q(1+\tov))\delta )\eqno(4.7)$$
It has to be noted that $Q(1+a)$ vanishes in the infinite volume limit.
Taking this limit simplifies the calculation, but obviously
prevents us from checking it using exact results at finite volume.

We can now manipulate the symbols introduced above
by just using their multiplication table and the rules to perform their
Fourier transform.
The multiplications appearing in Eqs. (3.13-16) can be performed as
a multiplication denoted $\star $ (respectively  $\ast $ in Fourier
transform)
in an abstract associative algebra where the basis
of the underlying vector space is given by the symbols introduced in Eq. (4.2)
(resp. Eq. (4.3)). Since in these two equations
$s$ can be understood as a continuous index,
this vector space is infinite dimensional. However, for calculations at
finite order, only finite dimensional subspaces are used.
The multiplication table for the ${\cal X}_s$ and $\delta$ is
$$\eqalign{&{\cal X}_s \star{\cal X}_t={\cal X}_{s+t} \cr
 & {\cal X}_s \star \delta =\delta \cr
& \delta \star \delta=\delta }\eqno(4.8)$$
Similarly for ${\cal K}_s $ and $\Delta$, we have
$$\eqalign{&{\cal K}_s \ast{\cal K}_t={\cal K}_{s+t}\cr
& {\cal K}_s \ast\Delta =2^{-ns\tov}\Delta \cr
& \Delta \ast \Delta=\Delta}\eqno(4.9)$$

A straightforward but tedious calculation yields the Fourier
transforms
$$\eqalign{&\widehat{{\cal X}_s}=(G(s))^{-1}({\cal K}_{(s-1){D \over 2}}
-2^{s-1}{\cal K}_0 )+2^n Q(s)\Delta \cr
&\widehat{{\cal X}_0}=2^n \Delta\cr
&\widehat{\delta}=1 }\eqno(4.10)$$
and their inverse
$$\eqalign{&\widehat{{\cal K}_a}=G(1+a)({\cal X}_{1+a}-Q(1+a){\cal X}_0)
+2^a\delta \cr
&\widehat{{\cal K}_0}=\delta \cr
&\widehat{\Delta }=2^{-n} {\cal X}_0 }\eqno(4.11)$$
The basic ingredient in this calculation is the identity (3.12).
However, a more organized calculation can be performed using
general results given in Ref. [\taibleson] (see next section).
With these results, one can calculate $\widehat{{\bf K}} (k)$ appearing in
Eq. (3.10)
$$\eqalign{\widehat{{\bf K}} (k)\ = &
P(1+{1\over D})(G(1+{1\over D}))^{-1}\ .\cr
& ({\cal K}_1+ (-{1\over 2}P(1+{1\over D})+Q(1+{1\over D})G(1+{1\over D}))
{\cal K}_0) \ .} \eqno(4.12)$$
This results can also be used for Feynman diagram calculations.

Finally, we need to evaluate all these functions when their argument
is 0. This again can be accomplished in a purely algebraic fashion,
by multiplying by $\Delta $ (resp. $\delta $) when an odd (resp. even )
number of Fourier transform has been performed and retaining the coefficient
of $\Delta $ (resp. $\delta $).

Using these results we find (recalling
the definition of $c$ given in (2.2))
\def\cf{{c\over 4}}
\def\ct{{c\over 2}}
\def\ce{{{c^2}\over 8}}
$$b_{1,n}=(1-\cf )^{-1} (\cf (1-(\ct)^n )
(1-\ct)^{-1} - (2^n -1)(\cf )^{n+1})\eqno(4.12)$$
and
$$\eqalign{b_{2,n}=&(b_{(1,n)})^2 - (1-\cf )^{-2}
[(\cf)^2 (1-(\ce)^n ) (1-\ce)^{-1} \cr
&-2 (\cf )^{n+2}(1-(\ct)^n ) (1-\ct)^{-1}
- (2^n -1)(\cf )^{2(n+1)}]}
\eqno(4.13)$$

The next two coefficients are more involved and
writing them explicitly might not be the most efficient way
to communicate the information to the reader.
Instead, we give in Appendix 1 a
Mathematica program which calculates the first four coefficients and allows
further
symbolic or numerical manipulations.
This program has been tested in various ways. The most convincing check
is the precise comparison between the numerical values obtained
by replacing $D$ and $n$ in the expression of the coefficients
and the exact values obtained by introducing a high temperature
expansion in the recursive integration used in Ref. [\num].
A program performing this calculation is given in Appendix 2.

Before closing this section, let us
recall that in order to obtain a physically interesting model,\refmark\dyson
it is essential to keep $b_{1,n}$ finite in the limit where $n$ becomes
infinite. Otherwise, flipping one spin in one of the configuration
where all the spins are aligned causes an infinite change in energy.
this requirement imposes $|c|<2$. If in addition we require ferromagnetic
couplings, we see from Eq. (2.4) that $c$ must be real,
strictly positive and less
than four. Combining the two requirements yields $0<c<2$ which
incidently corresponds to $D$ real, strictly positive and finite.

\chapter{Remarks Concerning the 2-adic Formulation of the Hierarchical Model}

Some of the results presented in the previous sections can
derived more elegantly if we had started immediately with the reformulation
of Ref. [\missarov ]. In this section, we briefly explain how this can
be done. Unlike the previous sections, this one assumes some
familiarity with the $p$-adic numbers.

The function $2^{v(x,y)}$ introduced in section 2, is a regularized
version of the 2-adic distance. Namely,
$$2^{v(x,y)} = \cases{  |x-y|_2 \ {\rm if}\ |x-y|_2 > 1  \cr
                     1\ {\rm if} \ |x-y|_2 \leq 1 } \eqno(5.1)$$
Using this, the invariance of $v(x,y)$ given in Eq. (2.10)
follows easily.

Similarly, $N(k)$ introduced in section 4, is a regularized version
of the 2-adic norm restricted to the 2-adic integers.
$$N(k)=\cases{ |k|_2 \  {\rm if} \ 1\geq |k|_2 > 2^{-n}  \cr
 2^{-n}\ {\rm if } \ |k|_2 \leq 2^{-n}} \ . \eqno (5.2)$$
%It takes
%its usual values unless $|k|_2 \leq 2^{-n}$, in which case it takes
%the value $2^{-n}$.
Given this, one can use the results of Ref. [\taibleson]
to calculate the Fourier transforms. In particular, the function $G(1+a)$
is the gamma function associated to a set of multiplicative characters.

\chapter{The Poles in the $c$-Complex Plane, in the Infinite Volume
Limit}

In the infinite volume limit, the expressions for the coefficients
simplify significantly. This limit will always be understood in this section
and we shall omit the reference to $n$. In taking this limit, we assume
$0<c<2$.
For instance, for the first coefficient,
we obtain
$$b_1=\sum_x{\bf K}_{0x}=
\widehat{{\bf K}}(0)={{2c} \over {(4-c)(2-c)}} . \eqno(6.1) $$
{}From Eq. (2.4), it is obvious that for any $m$, $b_m$ will have
a zero of order $m$ at $c=0$ and a pole of order $m$ at $c=4$.
Also, from the procedure used to calculate the coefficients, one
can see that $b_m$ will have a $(b_1)^m $ term (which would
be the only contribution in the gaussian case). This term
has a pole of order $m$ at $c=2$.
In order to get rid of these zeroes and poles, we shall
consider $b_m/(b_1^m)$ in the calculations below. In addition, with this
normalization, the quantities calculated have a random walk
interpretation when $0<c<2$. This is due to the fact that
${\bf K}_{xy}/\sum_z{\bf K}_{0z}$ can then be seen as the probability
for going from $x$ to $y$ in one step.

A quantity which is easy to calculate is the probability for visiting
no more than two sites, including the starting point, after $m$ step.
This quantity reads $\sum_x ({\bf K}_{0x})^m /(\sum_z{\bf K}_{0z})^m $
and appears in the calculation of $b_m/(b_1)^m $. After reexpressing
it in terms of the Fourier transform and using the techniques
developed in section 4, we obtain
$${{\widehat{{\bf K}^m }(0)} \over {(\widehat{{\bf K}}(0))^m}}=
{{{{\left( 2 - c \right) }^m}}\over {{2^{-1 + 2\,m}} - {c^m}}}\eqno(6.2)$$

Proceeding similarly for the other quantities entering in $b_3 /(b_1)^3$
and $b_4 /(b_1)^4$, we obtain
$$
{{\widehat{\widehat{{\bf K}}^3 }(0)} \over {(\widehat{{\bf K}}(0))^3}}=
{{6\,{{\left( -2 + c \right) }^3}\,{c^2}}\over
   {\left( 8 - {c^2} \right) \,\left( -16 + {c^3} \right) }},\eqno(6.3)$$
$$
{{\widehat{\widehat{{\bf K}}^4 }(0)} \over {(\widehat{{\bf K}}(0))^4}}=
{{{\left( -2 + c \right) }^4}{\left({-256\,}- {160\,{c^2}}
- {14\,{c^5}}+ {16\,{c^3}}\right)}
\over {\left( -8 + {c^2} \right) \,\left( -16 + {c^3} \right) \,
      \left( -32 + {c^4} \right) }}
\eqno(6.4)$$
and
$${{(\widehat{\widehat{{\bf K}^2}\wkn^2})(0)} \over \widehat{{\bf K}^4 }(0) }
={{{{\left( -2 + c \right) }^4}\,{\left({-128\,{c^2}}
- {64\,{c^3}}
+ {12\,{c^5}}\right) }}
\over {{{\left( -8 + {c^2} \right) }^2}\,\left( 8 + {c^2} \right) \,
      \left( -32 + {c^3} \right) }} \eqno(6.5)$$

Note that Eqs. (6.3) and (6.4) can be generalized to the calculation
of the probability for a return at the starting point after $m$ steps.
A detailed calculation shows that this quantity can be expressed as
$$
{{\widehat{\widehat{{\bf K}}^m }(0)} \over {(\widehat{{\bf K}}(0))^m}}=
\sum_{l=0}^m (-1)^l {m \choose l} {{(4-c)^l} \over {2^{l+1}-c^l}}\eqno(6.6)$$
This shows that unless unexpected cancelation occurs, $b_m$ will have
poles at $c=2^{1+1}, 2^{1+1/2},....,2^{1+1/m}$ or in other words at
$D=-2,-4,......,-2m$. On the other hand, the quantity appearing
in Eq. (6.6) has a zero of order $m$ at $c=2$. This is not obvious
from the r.h.s, however it can be seen by expanding the denominator
about 2 and summing over $l$, that this is the case for each term of
the resulting expression. The location of these poles
plays an important role if
one tries\refmark\app
to use the $1/D$-expansion to calculate the corrections to
the gaussian contribution.

\chapter{Conclusions}
We have shown that the symmetries of the hierarchical model can used
extensively to calculate the high temperature expansion of the magnetic
susceptibility. Up to order 4, the calculation can be performed using
algebraic methods which can be implemented by a computer program.
Proceeding this way, we  obtain analytical expressions where $D$ and $n$
are arbitrary. Substituting numerical values into these formulas
provides very precise numerical comparisons with other numerical methods.
The results of the numerical methods at low $n$ are very reliable and can
be used to test the analytical results.
If the analytical results pass this test, they can provide
very accurate results at
large $n$ and, in turn, provide
a test for the  numerical stability of the numerical method.

The simplicity of the calculations presented here
relies on the fact that at the end,
the arguments of the functions are set to zero. It will not be necessarily
the case for higher order calculations and it is conceivable that
some complicated sums will need to be performed numerically.
Nevertheless, it is clear that the methods proposed here provide
a drastic simplification when compared to a straightforward evaluation
of the sums as they appear in Eqs. (3.6-9).

In the infinite volume limit, one obtains more compact formulas. The
coefficients can be written as ratios of polynomials in $c$.
The location of the poles at each order seems to obey some regularity.
For the contributions which are calculable for arbitrary order, we see
that when the order increases, some of the poles get
closer to $c=2$ and $c=4$. This fact has to be kept in mind
if one attempts
to calculate the
corrections to the gaussian approximation using a $1/D$ expansion.

\Appendix{ 1: Calculation of the $b_{n,m}$}
This is a Mathematica program
which calculates the first four coefficients. The explanations
between (* ...*) play no role . If this program is stored in a file named
$bees.math$ for instance, then it can introduced in a Mathematica session
with the command $<<bees.math$. When all the instructions are completed
(it takes a few minutes),
you will see $completed$. You can then use $b3$ for $b_3$ etc...
In order to check numerical values with the program given in Appendix 2,
you can type for instance $ N[\{b1,b2,b3,b4\}/.\{n\rightarrow5,di
\rightarrow 3\},10]$, you
should then obtain
$\{1.075127131, 0.8953433348, 0.6726561277, 0.4555210038\}$.
This program is appended to the tex file.

\Appendix{ 2: Checking the Previous Calculation at Finite Volume}
This is a Mathematica program which calculates numerically the coefficients
up to a given order for $2^n$ sites. It is convenient to take $n<8$.
The numerical values of $n$ and $D=di$ can be changed. The choice of values
below, allows to check the numbers mentioned in Appendix 1. After
entering the program in a Mathematica session, you get (after a few minutes)
$1. + 1.075127131 B + 0.8953433348 B^2  + 0.6726561277 B^3  +
0.4555210038 B^4$.
This program is appended to the tex file.

\ack
I would like to thank V.G.J. Rodgers for suggestions concerning the
computer program mentioned at the end of the article and G. Ordaz
for providing
independent checks of the results presented in section 3 and 4.
This work has been completed during my stay at Brookhaven National
Laboratory, I would like to thank the theory group for its
hospitality.
\refout
\vfill
\eject
\end